\documentclass{svjour3}                     % onecolumn (standard format)
\usepackage{graphicx}
\usepackage{bm}

\def\prd{Physical Review D}
\begin{document}

\title{Critical Collapse of  Scalar Fields  Beyond Axisymmetry}
\author{James Healy$^{1,2}$ and Pablo Laguna$^1$}
\institute{$^1$Center for Relativistic Astrophysics and
School of Physics,
Georgia Institute of Technology, Atlanta, GA 30332\\
$^2$Center for Computational Relativity and Gravitation,
School of Mathematical Sciences,
Rochester Institute of Technology, 85 Lomb Memorial Drive, Rochester,
 New York 14623}

\date{Received: date / Accepted: date}

\maketitle

\begin{abstract}
We investigate non-spherically symmetric, scalar field collapse of a family of initial data consisting of a spherically symmetric profile with a deformation proportional to the real part of the spherical harmonic $Y_{21}(\theta,\varphi)$. Independent of the strength of the anisotropy in the data, we find that supercritical collapse yields a black hole mass scaling $M_h \propto (p-p^*)^\gamma$ with $\gamma \approx 0.37$, a value remarkably close to the critical exponent obtained by Choptuik in his pioneering study in spherical symmetry. We also find hints of discrete self-similarity. However, the collapse experiments are not sufficiently close to the critical solution to unequivocally claim 
that the detected periodicity is from critical collapse echoing.
\end{abstract}

%%%%%%%%%%%%%%%%%%%%%%%%%%%%%%%%%%%%%%%%%%%%%%%%%%%%%%%%%%%%%%%%%%%%%%%%%%%%
\section{Introduction} Without a doubt, one of the triumphs of 
numerical relativity occurred two decades ago, when Choptuik discovered 
universality, scale invariance and power-law scaling at the threshold 
of black hole formation. This remarkable body of work is commonly known as critical phenomena 
in gravitational collapse~\cite{1993PhRvL..70....9C}. Computationally, Choptuik's work was also revolutionary because it was the first instance in which adaptive mesh refinement techniques were used in numerical relativity. Without mesh refinement adaptivity, one could argue that unveiling  gravitational collapse criticality in such an exquisite detail would have been an insurmountable enterprise.

Naturally, early follow-up studies of  critical collapse focused on  configurations similar to those used by Choptuik, i.e., the spherically symmetric collapse of a scalar field. Subsequent studies showed that criticality is also found in the gravitational collapse of complex scalar fields, fluids and gravitational waves  (see \cite{2007LRR....10....5G} for a review). 

In broad terms, given a family of initial data parameterized by an attribute $p$,  critical collapse pertains to the threshold value $p^*$ such that $p > p^*$ data yield black hole formation while $p <p^*$ data disperse to infinity. 
Criticality surfaces in the limit $p \rightarrow p^*$, when the evolved data approach a critical solution. The solution is universal, characterized by an {\it echoing} parameter $\Delta$.
Furthermore, for collapses yielding black hole formation, i.e. supercritical cases, the black hole mass $M_h$ scales as $M_h \propto (p-p^*)^\gamma$, with $\gamma$ a universal {\it scaling} exponent whose value is independent of the family of initial data. Choptuik found that in the spherically symmetric massless scalar field collapse, $\gamma = 0.37$ and $\Delta = 3.4$.

Non-spherically symmetric critical collapse has not received as much attention. The main reason is likely  the computational cost imposed by 2D or 3D simulations as one approaches the $p \rightarrow p^*$ limit. This is particularly acute if the aim is to identify echoing. The first non-spherically symmetric study  was carried out by Abrahams and Evans~\cite{1993PhRvL..70.2980A} and focused on
axisymmetric vacuum gravitational collapse.  Abrahams and Evans found $\gamma \approx  0.38$, consistent with the value observed by Choptuik. The echoing, on the other hand, was found to be $\Delta \approx 0.6$, quite different from the value in spherically symmetric, scalar collapse.  
Collapse of non-linear gravitational waves has been also recently reconsidered by Hilditch et al.~\cite{2013PhRvD..88j3009H}. 
Also recently, Sorkin~\cite{2011CQGra..28b5011S} re-examined axisymmetric vacuum gravitational collapse in the subcritical regime, i.e. no black hole formation. His study found that  maximal values of curvature invariants also exhibit a power-law scaling with $\gamma \approx  0.38$. 

Regarding scalar fields, Mart\'in-Garc\'ia and Gundlach~\cite{1999PhRvD..59f4031M} studied non-spherical linear perturbations of the spherically symmetric critical solution. They found that all non-spherical perturbations decay, thus reducing the dynamics to that of the spherical symmetric case. Furthermore, they conjectured that non-linear perturbations will behave similarly. On the other hand, a non-linear axisymmetric study by Choptuik et al.~\cite{2003PhRvD..68d4007C} suggests the existence of a slowly growing, non-spherical mode about the spherically symmetric critical solution. However,  they state that their results are not conclusive since they could not rule out  that the observed growing mode was of pure numerical origin.

Our objective is the critical collapse of a scalar field beyond axisymmetry. We concentrate on the case of a massless scalar field and consider anisotropic collapsing scalar configurations. The family of initial data consists of a parameterized deformation of a spherically symmetric configuration. We find that, independent of the degree of anisotropy, the collapse yields the spherically symmetric, critical scaling exponent $\gamma \approx 0.37$. In addition, the simulations producing the smallest black hole masses show some hints of  discrete self-similarity. That is, we detect periodicity that resembles echoing with parameter values $\Delta \sim 3.1$. Measuring $\Delta$ necessitates looking at solutions near $p \rightarrow p^*$. Simulations in this limit were not computationally feasible in the present study. Thus, we were not able to unequivocally claim that the observed periodicity is the result of critical collapse echoing.

The simulations for this work were obtained with the \textsc{Maya} numerical relativity
code of our group~\cite{Bode:2009mt,Husa:2004ip,2011arXiv1111.3344L}. The \textsc{Maya} code is primarily used for compact object binary simulations. For the present work, the code was modified to include scalar field sources. The code is based on the BSSN formulation and the moving puncture gauge~\cite{2006PhRvL..96k1101C,2006PhRvL..96k1102B}. 

%%%%%%%%%%%%%%%%%%%%%%%%%%%%%%%%%%%%%%%%%%%%%%%%%%%%%%%%%%%%%%%%%%%%%%%%%%%%
\section{Initial Data Family}  

We considered the following family of scalar field initial data:
\begin{equation}
\Psi(r,\theta,\varphi) = \psi(r)[1 +\epsilon \,{\rm Re}\,( Y_{21}(\theta,\varphi))]
\end{equation}
where
\begin{equation}
\psi(r) = p \left[ 1 - \tanh\left( \frac{r-r_0}{\sigma} \right) \right] \label{eq:tanh}
\end{equation}
and
\begin{equation}
Y_{21}(\theta,\varphi) = - \sqrt{ \frac{15}{8\pi} }e^{i\,\varphi}\sin{\theta}\cos{\theta}\,.
\end{equation}
That is, the scalar field consists of a flat-top, spherically symmetric profile $\psi(r)$ with a deformation proportional to the real part of 
the spherical harmonic $Y_{21}(\theta,\varphi)$. The parameter $\epsilon$ controls the strength of the anisotropy, with $\epsilon = 0$ the spherically symmetric case. For each value of $\epsilon$, we carried out critical collapse experiments in which the amplitude parameter $p$ plays the role of the attribute determining the threshold of black hole formation. For the present work, we set in Eq.~(\ref{eq:tanh}), $\sigma = 2\,M$ and $r_0 = 10\,M$ with $M$ an arbitrary mass scale in the code. We assume the initial data to be time-symmetric, i.e. $\partial_t \Psi = 0$, and use units in which $G = c = 1$.  

The computational domain involves a grid structure with 10 levels of refinements. The finest mesh has a grid spacing of $M/94$ and size extent of $0.94\,M$.  The outer boundary is located at $240\,M$. For supercritical data,  black hole formation typically occurs at $t \sim 20\,M$. We observe that after $t >40\,M$ the black hole mass $M_h$ stabilizes, with $M_h$ measured from the area of the apparent horizon.   To investigate the black hole mass scaling, $M_h \propto (p-p^*)^\gamma$, we take the value of the black hole mass as that at times  $t \sim 150\,M$ from the beginning of the simulation. At this point in time, most of the scalar field has been either dispersed or accreted by the black hole; in other words, to a good approximation, we can consider having a black hole in isolation. Our operational definition of isolated black hole is when the accretion time-scale $M_h/\dot M_h >  10^4\,M$.  We should also point out that at times $t \sim 150\,M$, when we measure  the black hole mass, it is still early enough in the evolution to ensure that any disturbances from reflections at the outer boundary have not had enough time to reach the newly formed black hole and thus affect its mass measurement. 

%%%%%%%%%%%%%%%%%%%%%%%%%%%%%%%%%%%%%%%%%%%%%%%%%%%%%%%%%%%%%%%%%%%%%%%%%%%%
\section{Power-law Scaling}  

We present results from critical collapse experiments with anisotropy parameter values $\epsilon = \lbrace 0,0.25, 0.5,1\rbrace$. For each value of $\epsilon$, our first task was to identify a suitable starting value of $p$ that allows one to identify the $M_h \propto (p-p^*)^\gamma$ scaling. Not surprisingly, if one starts the sequence at $p \gg p^*$, a single power-law black hole mass scaling is not found. By experimentation, we found that a good starting value for $p$ is one resulting in a black hole with mass $M_h \sim 0.55\,M$. This translates into initial values  $p= \lbrace 0.0607, 0.0606, 0.0606, 0.0592\rbrace$ for $\epsilon = \lbrace 0,0.25, 0.5,1\rbrace$, respectively. With those initial amplitudes, we march toward $p^*$ in steps of $\sim 10^{-4}$ until we reached a black hole mass $M_h \sim 0.18\,M$. This is the smallest black hole mass that we are able to accurately resolve with the grid setup outlined in the previous section. There are no serious impediments reaching smaller black hole masses other than adding finer mesh refinements to our computational grid. We did precisely  that in the $\epsilon = 0$ case, and we were able to obtain black holes with masses as small as $10^{-3}\,M$. We found, however, that adding  smaller masses did not  change significantly the values measured for the critical exponent $\gamma$ or for the threshold value $p^*$.

\begin{figure}
\centering
\vbox{
    \includegraphics[width=.70\linewidth,angle=270]{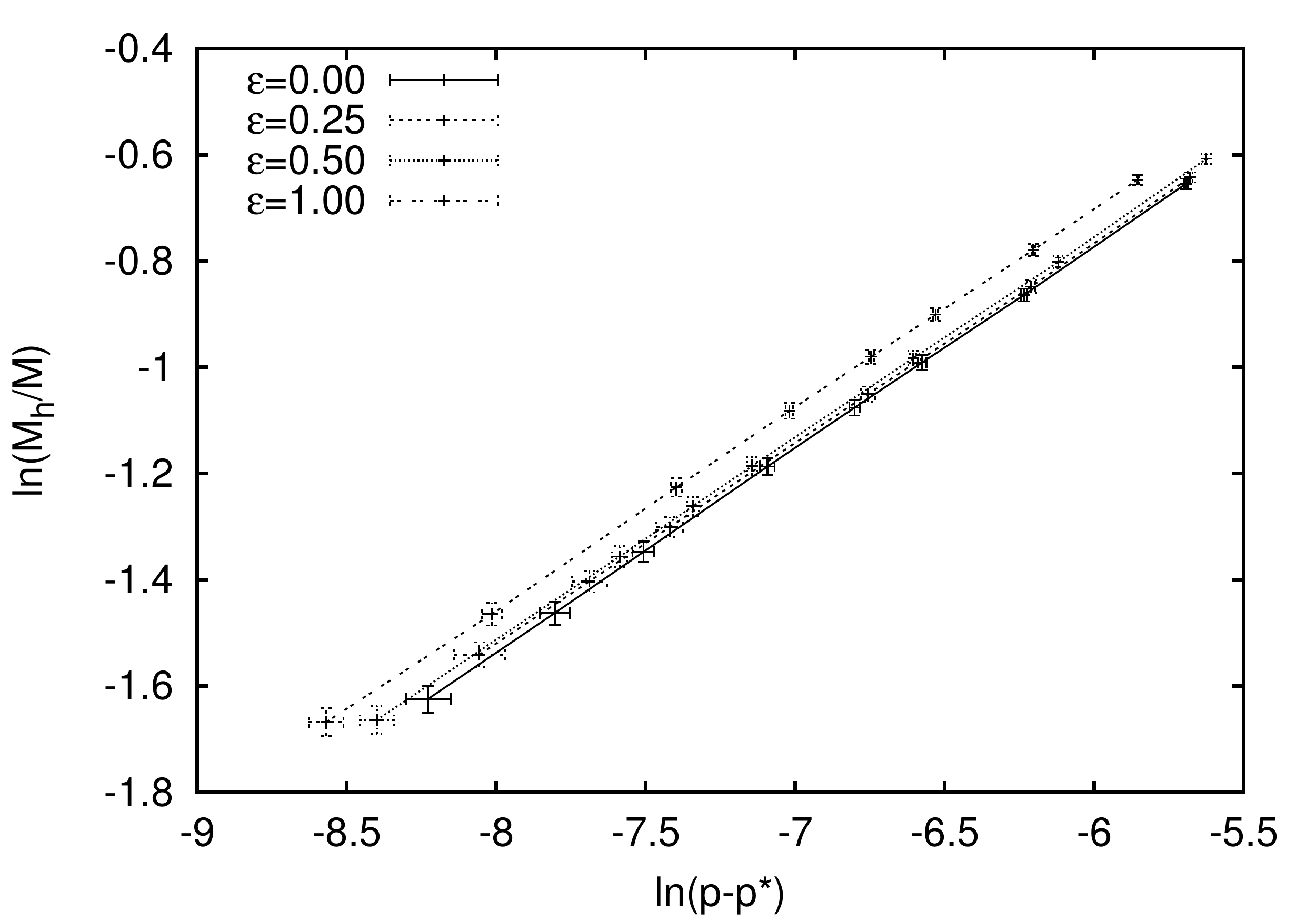}
}
\caption{Log-log plot of the black hole mass $M_h$ in units of $M$ as a function of $p-p^*$ for each value of $\epsilon$. Straight lines are least squares fits to black hole mass scaling $M_h/M = C (p- p^*)^{\gamma}$.}
\label{fig:critexp}
\end{figure}

Figure~\ref{fig:critexp} shows a log-log plot of black hole mass $M_h$ in units of $M$  versus $p-p^*$. The values of $p^*$ used for each plot are those reported in Table~\ref{tab:critexp}. The error bars in the black hole masses in Fig.~\ref{fig:critexp}  are obtained from a numerical convergence analysis using simulations with 
resolutions of $88^3$, $104^3$, and $120^3$ grid points.  
Not surprisingly, the largest errors are found  in the cases with the 
smallest final black hole. At the end of this section, we describe how the $p^*$ values are found as well as the error bars for $p-p^*$ in Fig.~\ref{fig:critexp}. 
The $M_h \propto (p-p^*)^\gamma$ scaling is evident from Fig.~\ref{fig:critexp}. A linear fitting to the data yields the critical exponent $\gamma$, which is reported in Table~\ref{tab:critexp}.  
Notice that, within our uncertainties, anisotropic collapse yields Choptuik's spherically symmetric critical exponent of $\gamma \approx 0.37$, suggesting the dominance of the spherically symmetric mode.

\begin{table}
  \begin{tabular}{|c|c|c|c|c|}
  \hline
$\epsilon$ & $\gamma$ & $p^*/10^{-2}$ & $C$ & $\Delta$\\
\hline
0.00 & $0.379\pm 0.008$ & $5.728\pm 0.002$& $4.50\pm 0.17$ & $3.08\pm0.16$\\
0.25 & $0.379\pm 0.010$ & $5.723\pm 0.003$& $4.51\pm 0.20$ & $3.26\pm0.34$\\
0.50 & $0.381\pm 0.005$ & $5.704\pm 0.001$& $4.62\pm 0.11$ & $3.09\pm0.32$\\
1.00 & $0.375\pm 0.006$ & $5.637\pm 0.001$& $4.71\pm 0.12$ & $3.27\pm0.38$\\
\hline
  \end{tabular}
\caption{Table with the critical exponent $\gamma$, threshold parameter $p^*$, fitting constant $C$ for the black hole mass scaling $M_h/M = C (p- p^*)^{\gamma}$, and echoing period $\Delta$. }
\label{tab:critexp}
\end{table}

\begin{figure*}[p]
\centering
\vbox{
    \hbox{
    \includegraphics[width=.35\linewidth,angle=270]{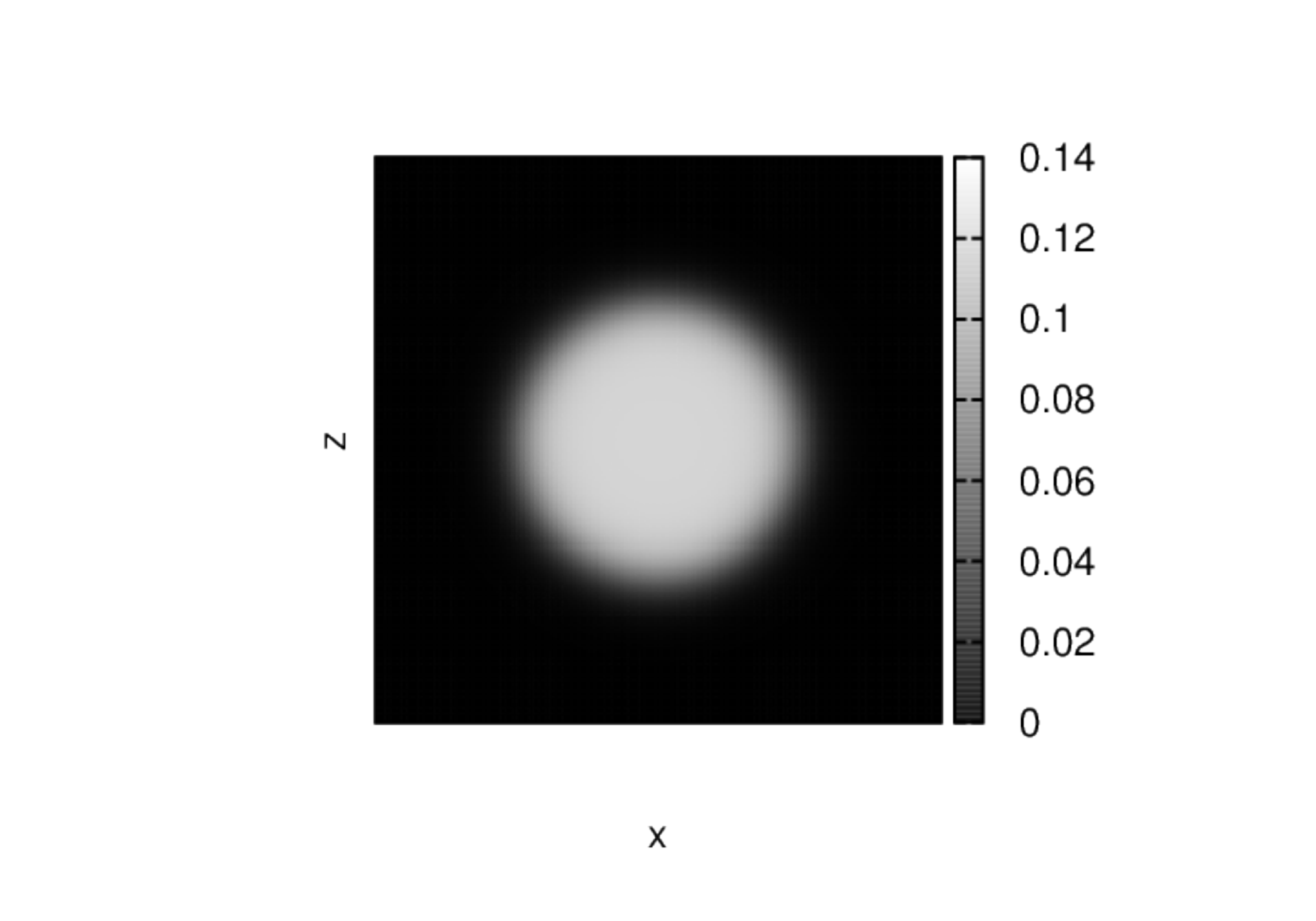}
    \includegraphics[width=.35\linewidth,angle=270]{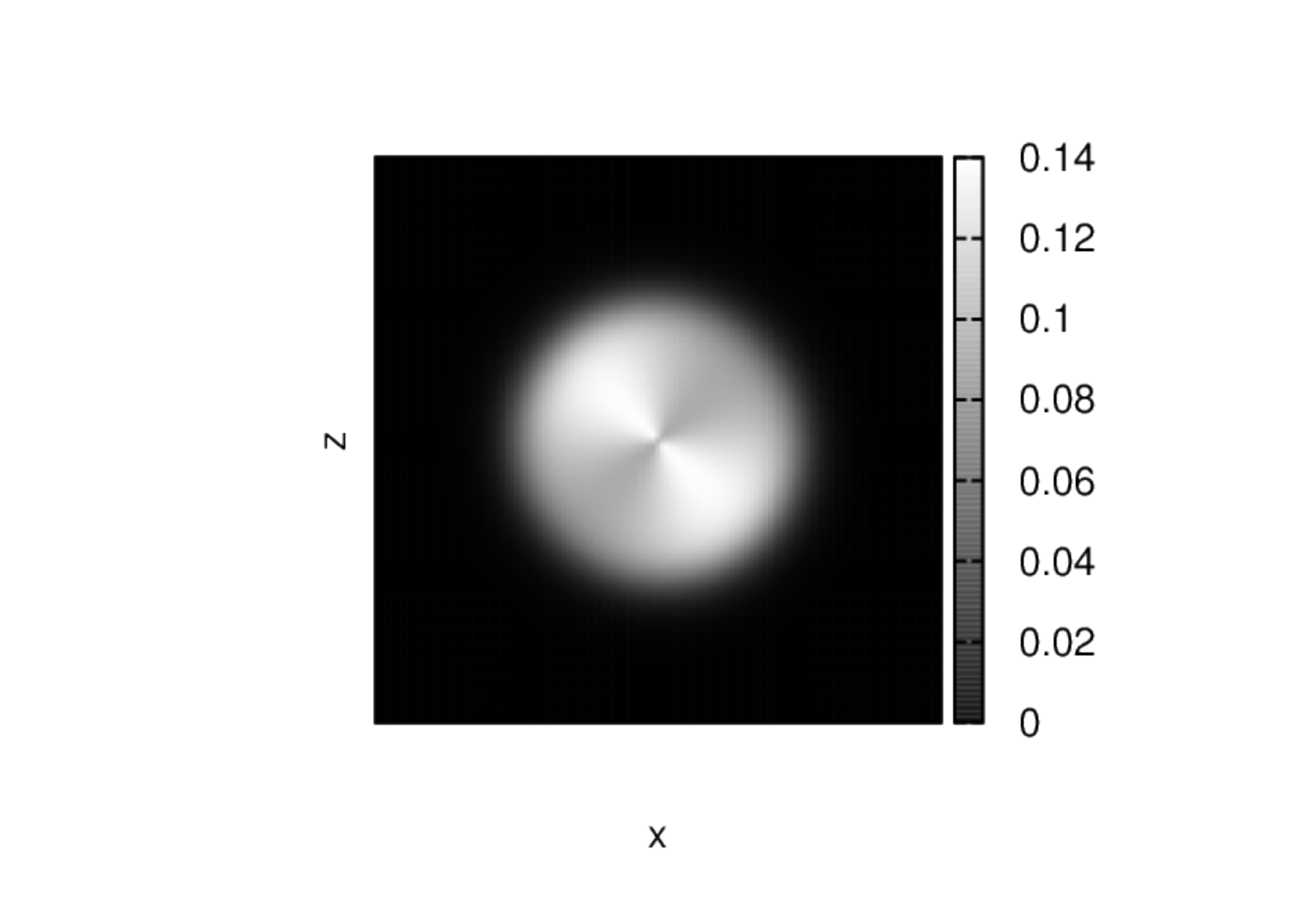}}
    \hbox{
    \includegraphics[width=.35\linewidth,angle=270]{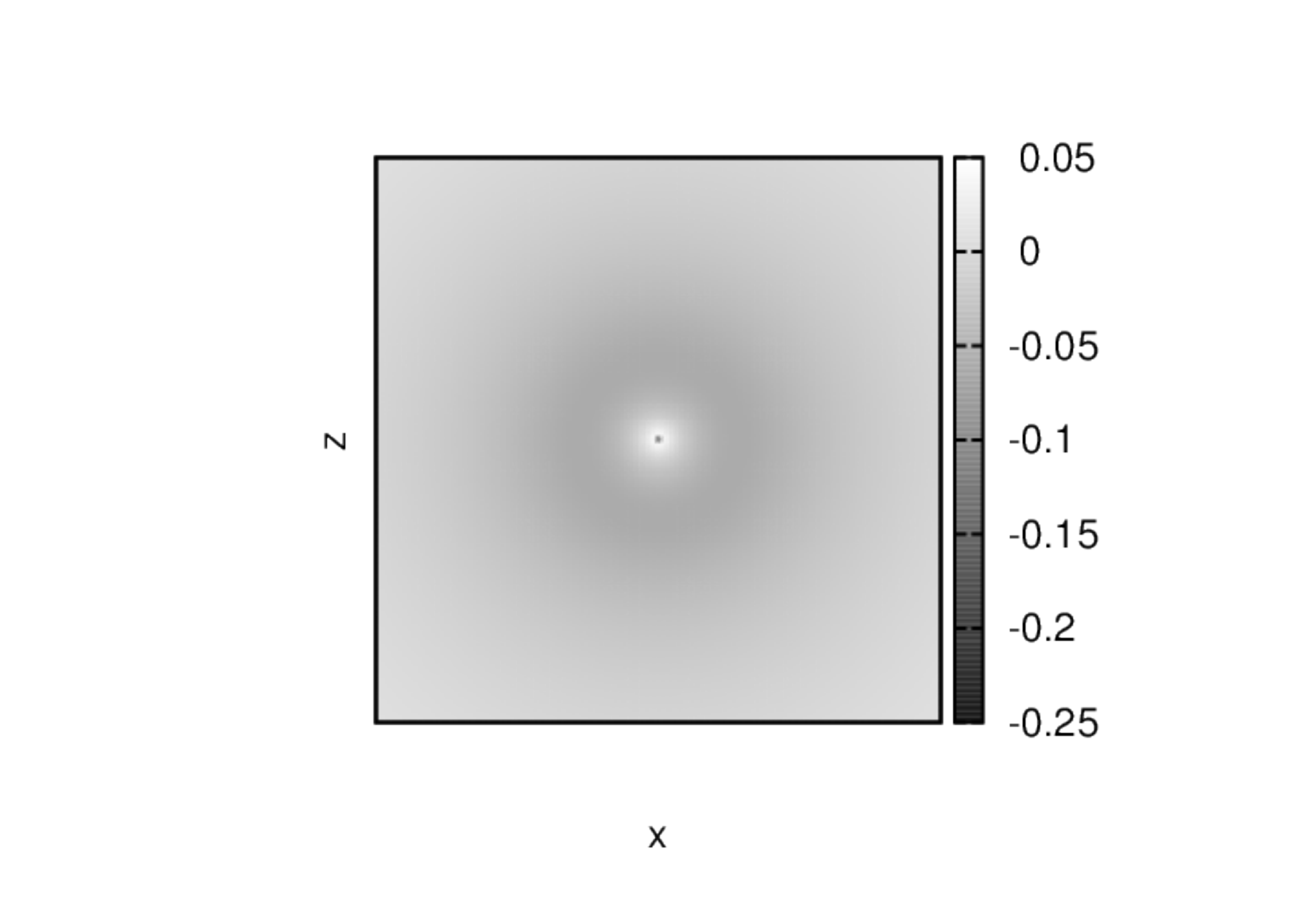}
    \includegraphics[width=.35\linewidth,angle=270]{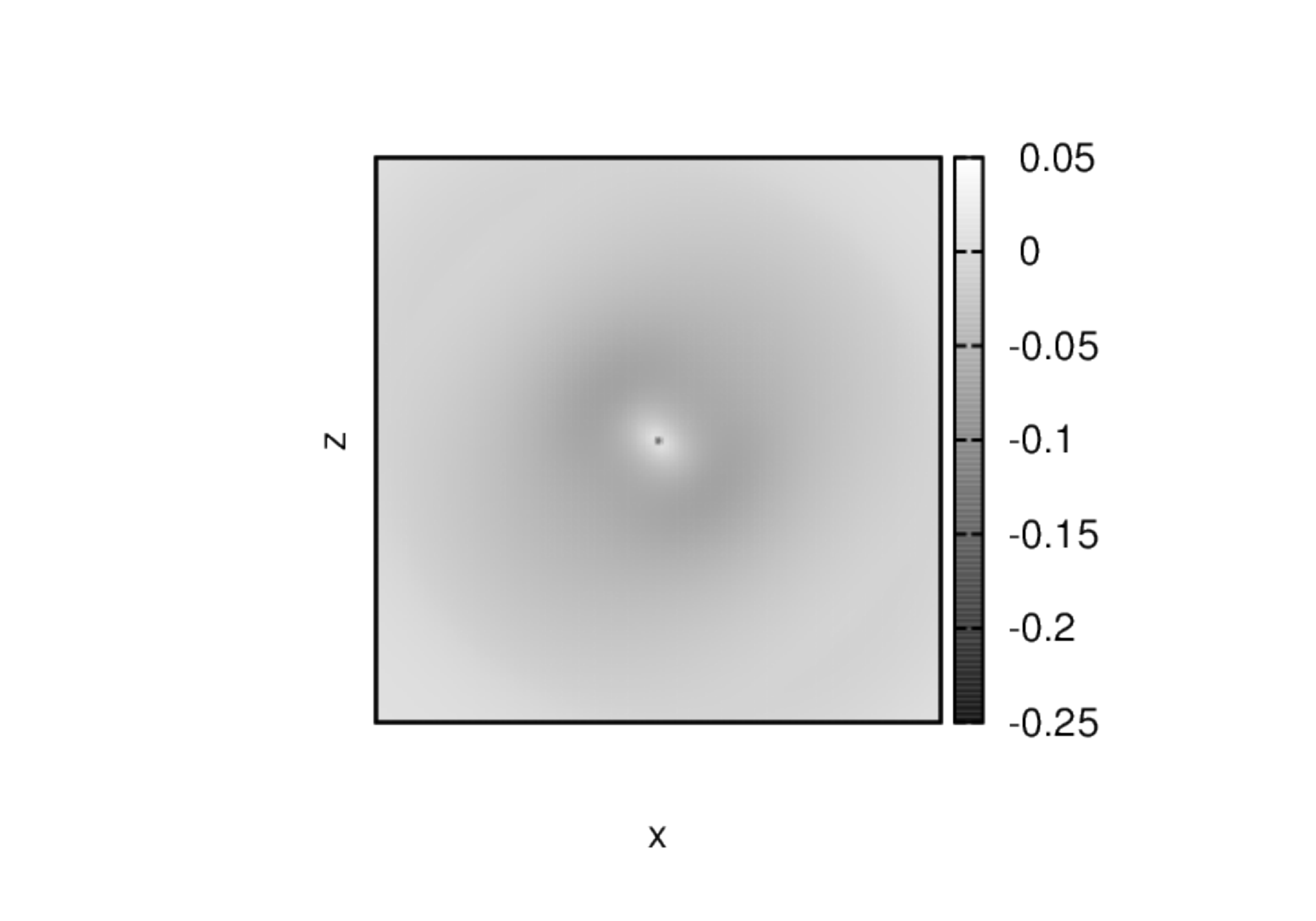}}
    \hbox{
    \includegraphics[width=.35\linewidth,angle=270]{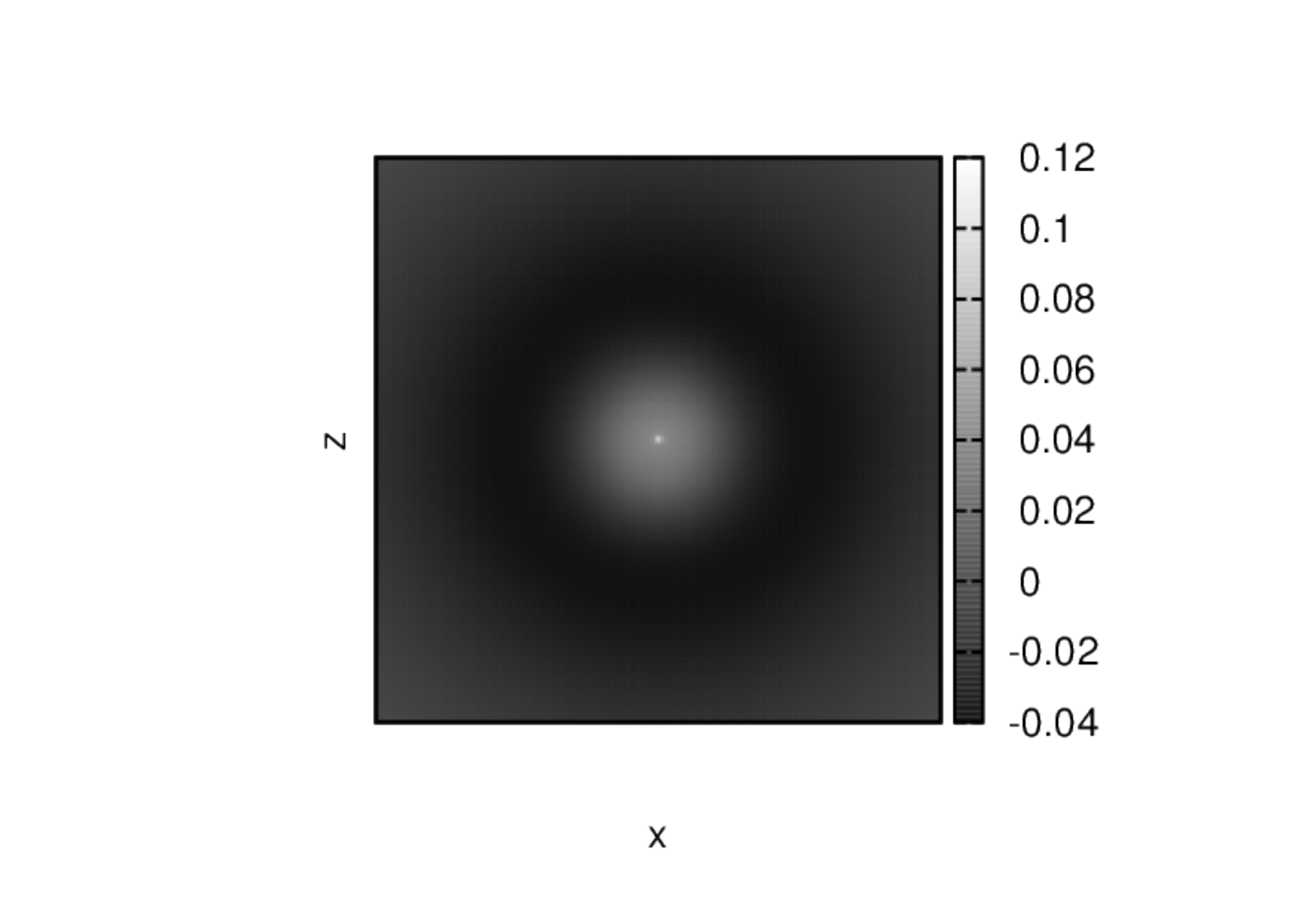}
    \includegraphics[width=.35\linewidth,angle=270]{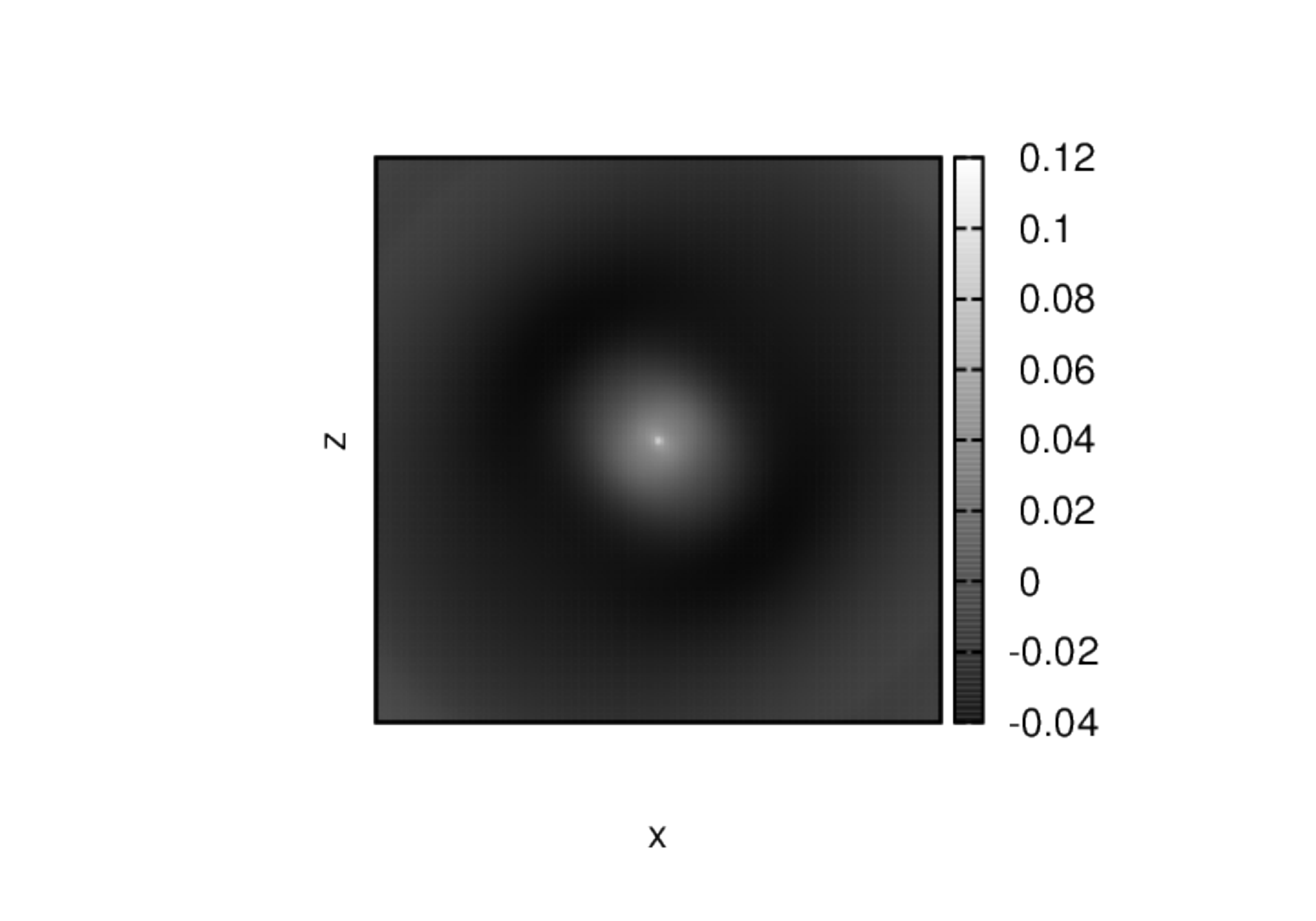}}
}
\caption{Snapshots of the scalar field in the $xz$-plane. Left column shows the $\epsilon=0.0$ case, right column shows the $\epsilon=0.5$ case.  Both have initial scalar field profiles given by Eq.~\ref{eq:tanh} with $p = 0.0584$.  
From top to bottom the snapshots correspond to times $t = 0\,M,\, 26.8\,M,$ and $34.6\,M$.  The middle row corresponds to the time when the first horizon is found.  The $x$ and $z$ ranges are between $-20\,M$ and $20\,M$.}
\label{fig:snapshots}
\end{figure*}

Further evidence for the dominance of the spherically symmetric mode can be found in Figure~\ref{fig:snapshots}. In this figure, we compare three simulation snapshots of the spherical symmetric case (left column) with the corresponding snapshots for the case with $\epsilon = 0.5$ (right column). In the top snapshots, taken at the start of the simulations, there are evident differences. The top left snapshot is clearly spherically symmetric, while the right one shows clear anisotropies.  The two snapshots in the middle in  Fig.~\ref{fig:snapshots} correspond to a time $t \sim 26.8\,M$, just after we find the apparent horizon of the black hole. Notice that there are still hints of anisotropies remaining in the snapshot at the right.  Finally, the bottom snapshots in Fig.~\ref{fig:snapshots} were taken at $t \sim 34.6\,M$. They are almost identical, indicating the disappearance of any traces of anisotropies.

\begin{figure}
\centering
\vbox{
   \includegraphics[width=.70\linewidth,angle=270]{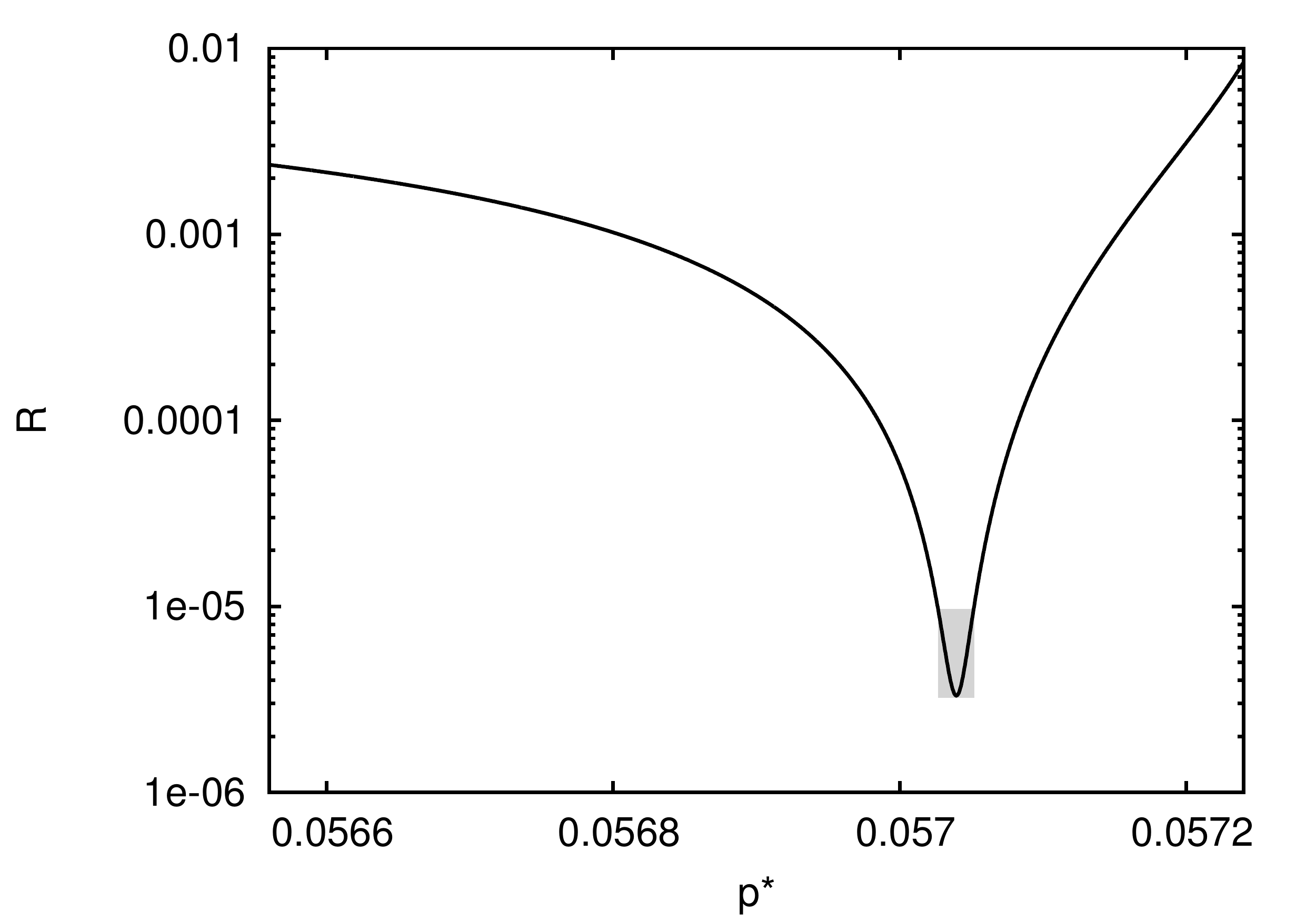}
}
\caption{Plot of the $L_2$-norm of the residuals ${\cal R}$ for the least-square fitting of the  black hole mass scaling $M_h/M = C (p- p^*)^{\gamma}$ as a function of the threshold parameter $p^*$. The grey area gives an estimate of the error in $p^*$.}
\label{fig:residuals}
\end{figure}

To produce the data as displayed in Fig.~\ref{fig:critexp}, one needs to find the threshold value $p^*$. 
To do so, we applied a non-linear, least squares fitting of the data $\lbrace M_h/M, p\rbrace$ to the
function  $M_h/M = \bar C (p-\bar p^*)^{\bar\gamma}$, with $\bar C, \,\bar p^*$ and $\bar \gamma$ the fitting parameters. With $\bar p^*$ at hand, 
we selected a range of possible values for $p^*$ in the interval $[0.95\,\bar p^*,  1.05\,\bar p^*]$.
For each value of $p^*$ within this interval,  we carried out a linear, least squares fitting of the data $\lbrace \ln{(M_h/M)}, \ln(p- p^*)\rbrace$ to the function   $\ln{(M_h/M)} = \gamma\,\ln{(p-p^*)} + \ln{(C)}$, with $\gamma$ and $C$ the fitting parameters. For each fitting, we calculated the $L_2$-norm of the residuals ${\cal R}(p^*)$. Figure~\ref{fig:residuals} shows ${\cal R}(p^*)$ for the case $\epsilon = 0.5$. The other cases have a similar behavior. Finally, we set the threshold parameter $p^*$ to be the value that minimizes ${\cal R}(p^*)$.

To estimate the error in $p^*$ reported in Table~\ref{tab:critexp}, we single out, among the residuals that went into constructing ${\cal R}(p^*)$, those that produced a linear fit passing within the error bars of the black hole mass with the largest error. In Fig.~\ref{fig:residuals} for instance, those are the residuals in the grey area. The width of this grey area gives an estimate of the error in $p^*$, and thus in $p-p^*$.
These are the errors used to estimate the errors for the critical exponent $\gamma$ and fitting constant $C$ reported in Table~\ref{tab:critexp} and displayed as the horizontal error bars in Fig.~\ref{fig:critexp}.  

%%%%%%%%%%%%%%%%%%%%%%%%%%%%%%%%%%%%%%%%%%%%%%%%%%%%%%%%%%%%%%%%%%%%%%%%%%%%
\section{Echoing}

One of the predictions of criticality for a single, non-interacting massless scalar field is discrete self-similarity. Specifically, the critical solution exhibits scale-periodicity in the sense that any dimensionless quantity $A$ in the system obeys 
\begin{equation}
A(\tau -\tau^*,r) = A(e^{n\Delta}(\tau -\tau^*),e^{n\Delta}\,r) 
\label{eq:dss}
\end{equation}
where $\tau$ is the central proper time defined by
\begin{equation}
\tau = \int_0^t \alpha(\tilde t,0)\,d\tilde t\,.
\end{equation} 
In Eq.~(\ref{eq:dss}), $\tau^*$ is the accumulation time of the self-similar solution, $n$ labels the ``echo'' number and $\Delta$ the echoing parameter. 

As indicated before, our simulations yield collapse solutions that are not near the $p \rightarrow p^*$ limit where echoing should in principle be evident. Nonetheless, we applied the same prescription as in Ref.~\cite{2007PhRvD..76l4014O} to investigate any hints of an echoing parameter $\Delta$. That is, we used the fact that the scalar field $\Psi$ or geometrical quantities such as the lapse function $\alpha$, satisfying Eq.~(\ref{eq:dss}), will have local extremal values at discrete central proper values $\tau_n$ such that
\begin{equation}
\label{eq:echo}
\tau_n - \tau^* = e^{n\Delta}(\tau_0 -\tau^*)\,,
\end{equation} 
where  $\tau_0$ is the starting time used to count the echoes. 

\begin{figure}
\centering
\vbox{
    \includegraphics[width=1.40\linewidth,angle=270]{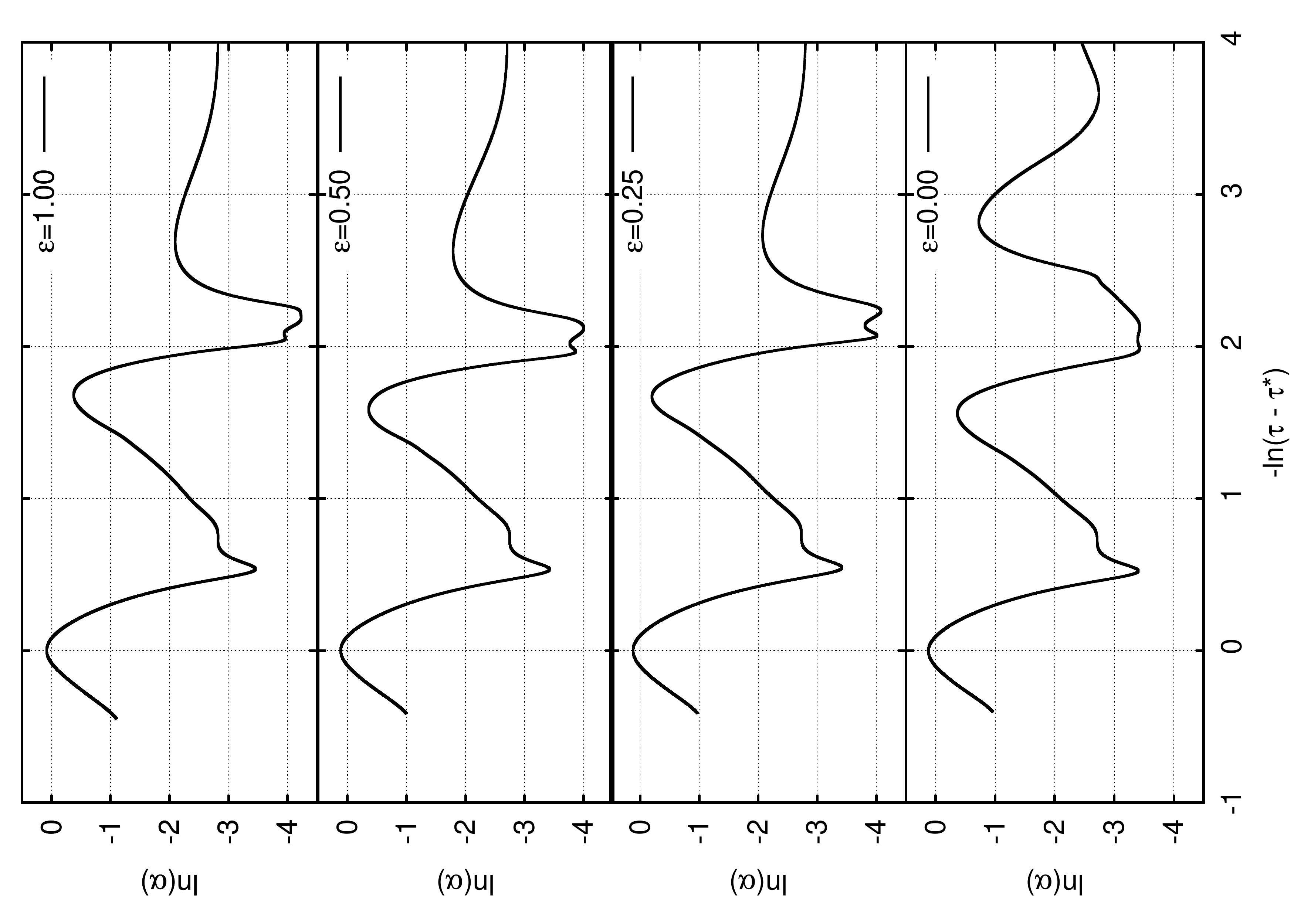}
}
\caption{Plot of the  $\ln(\alpha)$, with $\alpha$ the value of the lapse function at the origin,  as a function of $-\ln(\tau-\tau^*)$, with $\tau$ the
central proper time defined by Eq.~(\ref{eq:dss}).}
\label{fig:echo}
\end{figure}

\begin{figure}
\centering
\vbox{
    \includegraphics[width=1.40\linewidth,angle=270]{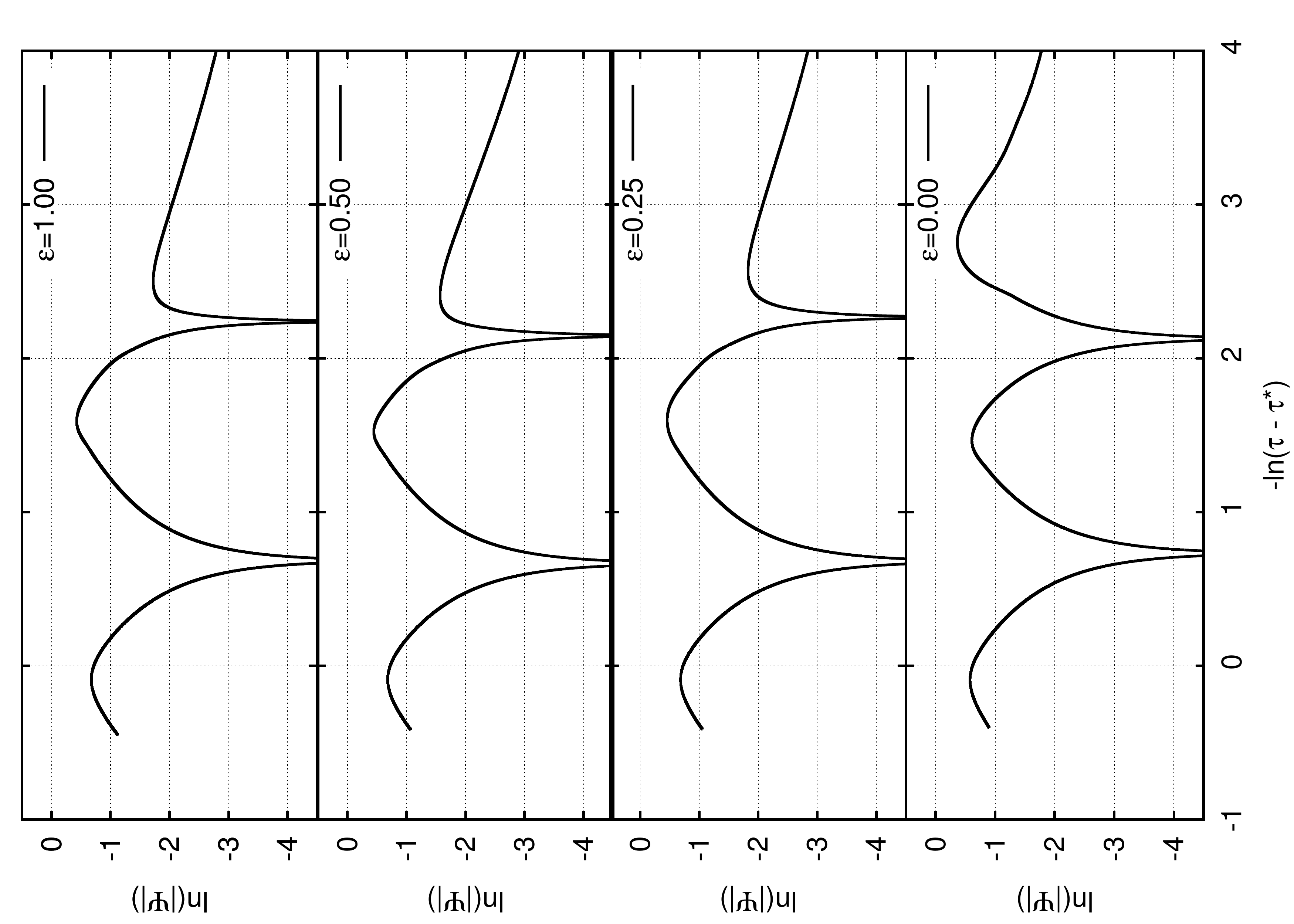}
}
\caption{Plot of the  $\ln(|\Psi|)$, with $\Psi$ the value of the scalar field at the origin,  as a function of $-\ln(\tau-\tau^*)$, with $\tau$ the
central proper time defined by Eq.~(\ref{eq:dss}).}
\label{fig:echo2}
\end{figure}

To measure $\Delta$ in Eq.~(\ref{eq:echo}), we carried out simulations with scalar field amplitudes
$p= \lbrace 0.05725, 0.05720, 0.05702, 0.05636\rbrace$ corresponding to $\epsilon = \lbrace 0,0.25, 0.5,1\rbrace$, respectively. We should point out that  for these scalar field amplitudes $p < p^*$. That is, we found it less challenging to identify something that resembles echoing in the subcritical regime. For each simulation, we monitored the value of the lapse function $\alpha$ at the origin and were able to identify at least three local extremal values. With those values at hand, a least squares fitting to Eq.~(\ref{eq:echo}) yielded both $\tau^*$ and the values of  $\Delta$ reported in Table~\ref{tab:critexp}. In Figures~\ref{fig:echo} and \ref{fig:echo2}, we show  the central values of $\ln(\alpha)$ and $\ln{(|\Psi|)}$ respectively as a function of $-\ln(\tau-\tau^*)$.The oscillatory behavior in both quantities is evident from these figures. To some degree it is reassuring that the value in Table~\ref{tab:critexp} of $\Delta = 3.08$ for the spherically symmetric case ($\epsilon = 0$) is close to the value of 3.4 found by Choptuik. However, given that our solutions are far from the $p \rightarrow p^*$ limit, we cannot with certainty affirm that the observed oscillation are indeed echoing, as hinted by Figs.~\ref{fig:echo} and \ref{fig:echo2}.

\begin{figure}
\centering
\vbox{
    \includegraphics[width=.70\linewidth,angle=270]{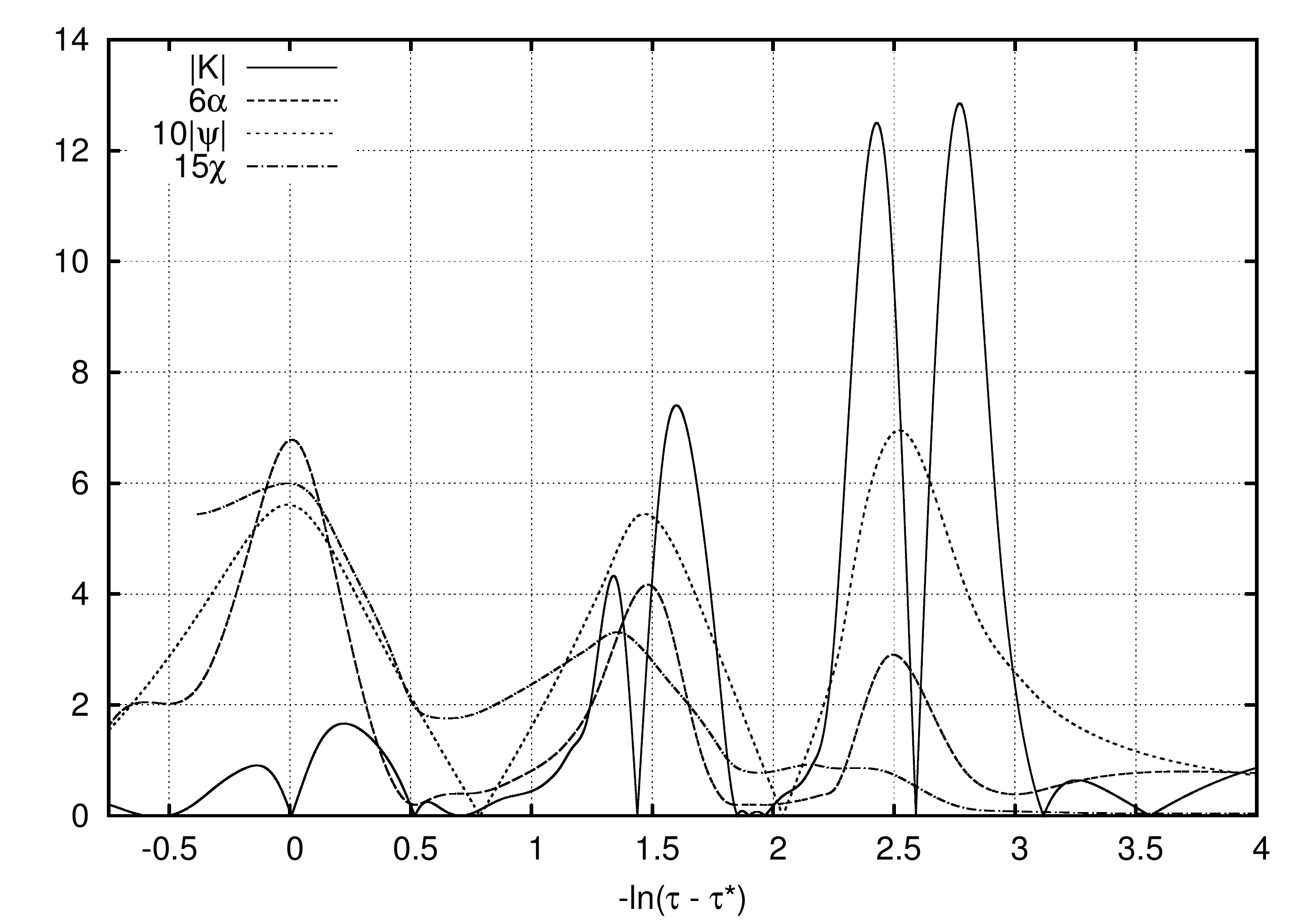}
    \includegraphics[width=.70\linewidth,angle=270]{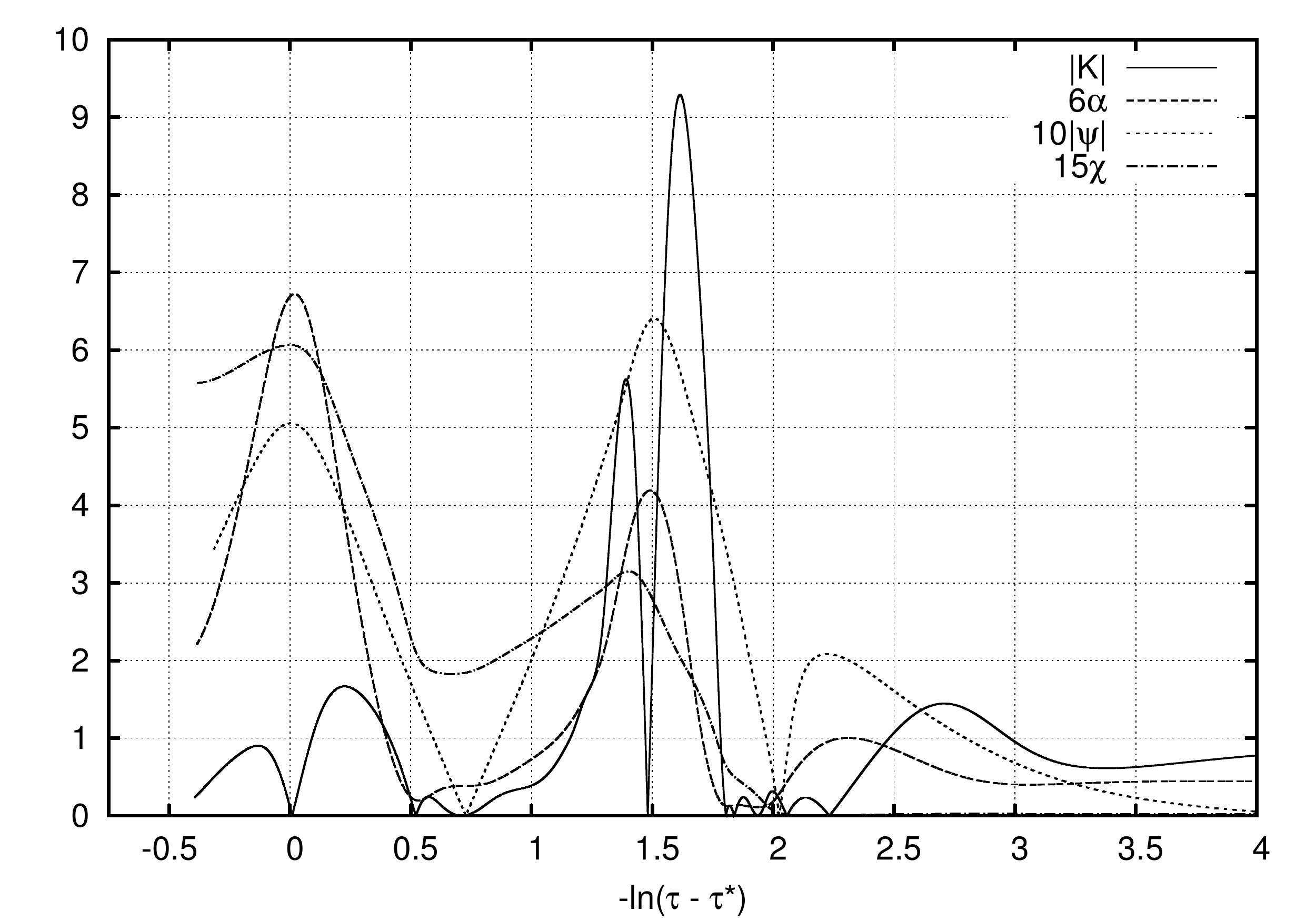}
}
\caption{Plot of the trace of the extrinsic curvature, $K$, lapse, $\alpha$, conformal factor, $\chi$, and scalar field, $\Psi$ as a function of  $-\ln(\tau-\tau^*)$ for $\epsilon=0.0$ in the top panel, and $\epsilon=0.5$ in the bottom panel.}
\label{fig:echoes}
\end{figure}

As pointed out by one of the referees, to demonstrate echoing one needs to use geometric invariants and observers that are not affected by the choice of coordinates. The lapse displayed in Fig.~\ref{fig:echo}  is certainly not a geometric invariant. An alternative is to show that the gauge conditions used, i.e. choice of lapse and shift, yield adapted coordinates where~\cite{2007LRR....10....5G}
\begin{equation}
g_{\mu\nu}(\tau,x^i) = e^{-2\,\tau}\tilde g_{\mu\nu}(\tau,x^i) \,,
\label{eq:DDS}
\end{equation}
 with $g_{\mu\nu}(\tau,x^i)$ periodic in $\tau$ with period $\Delta$. As mentioned in the introduction, in the present study, we use the moving puncture gauge conditions~\cite{2006PhRvL..96k1101C,2006PhRvL..96k1102B}. These conditions have been crucial for the success in binary black hole simulations. If the observed oscillatory behavior is indeed connected to discrete self-similarity, it would thus also suggest that the moving puncture gauge yields the adapted coordinate system required for Eq.~(\ref{eq:DDS}). Unfortunately, we were not able to formally verify  the emergence of adapted coordinates as the solution becomes oscillatory; however,  Figure~\ref{fig:echoes} supports this conjecture, where we plot, in addition to $\alpha$ and $\Psi$,  the trace of the extrinsic curvature $K$ and the exponent in the BSSN conformal factor $\chi$ (essentially the logarithm of the spatial metric determinant) as a function of   $-\ln(\tau-\tau^*)$. It is noticable  from this figure the oscillatory behavior in all metric related quantities. 

%%%%%%%%%%%%%%%%%%%%%%%%%%%%%%%%%%%%%%%%%%%%%%%%%%%%%%%%%%%%%%%%%%%%%%%%%%%%
\section{Gravitational Radiation}

\begin{figure}
\centering
\vbox{
    \includegraphics[width=.70\linewidth,angle=270]{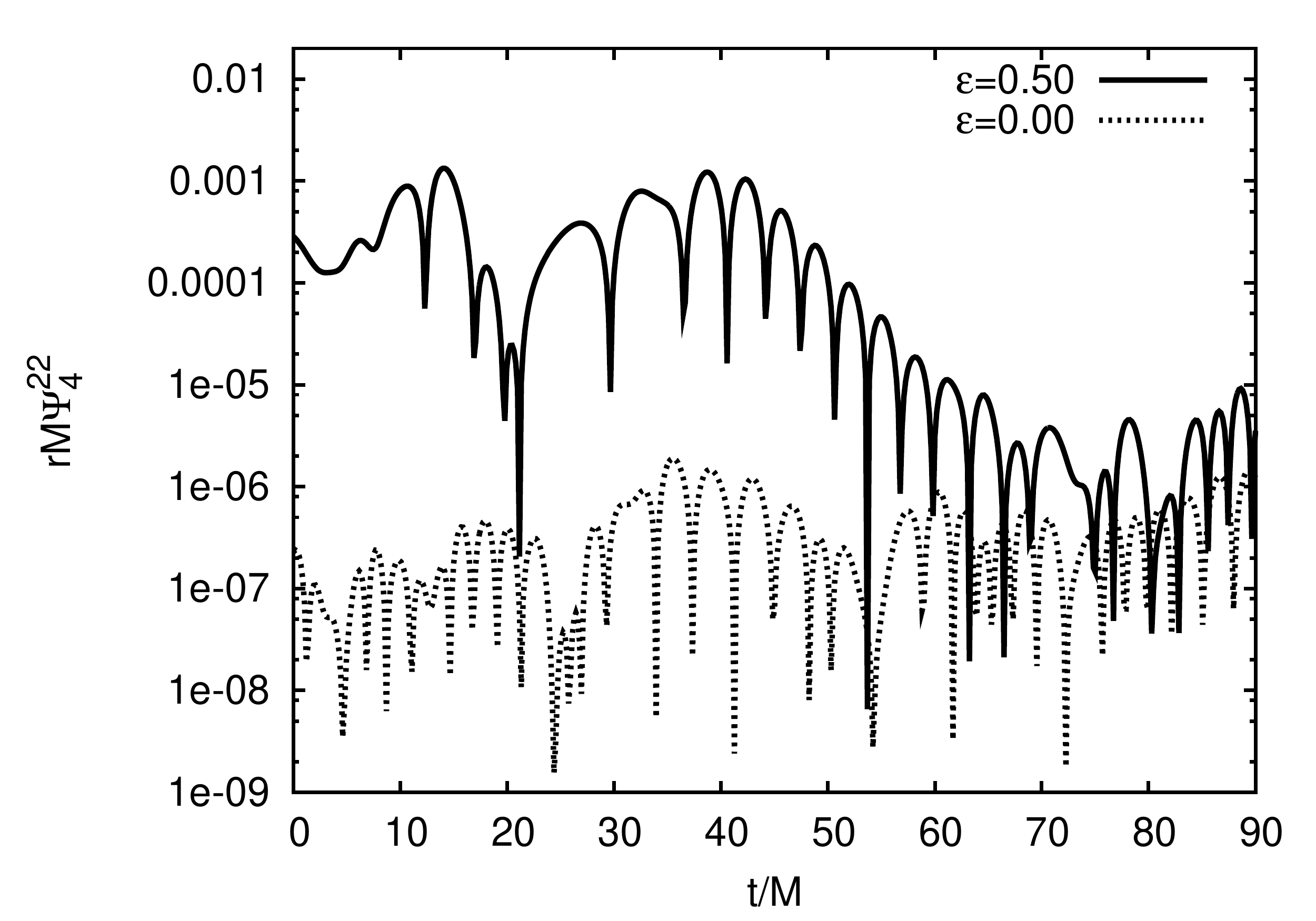}
    \includegraphics[width=.70\linewidth,angle=270]{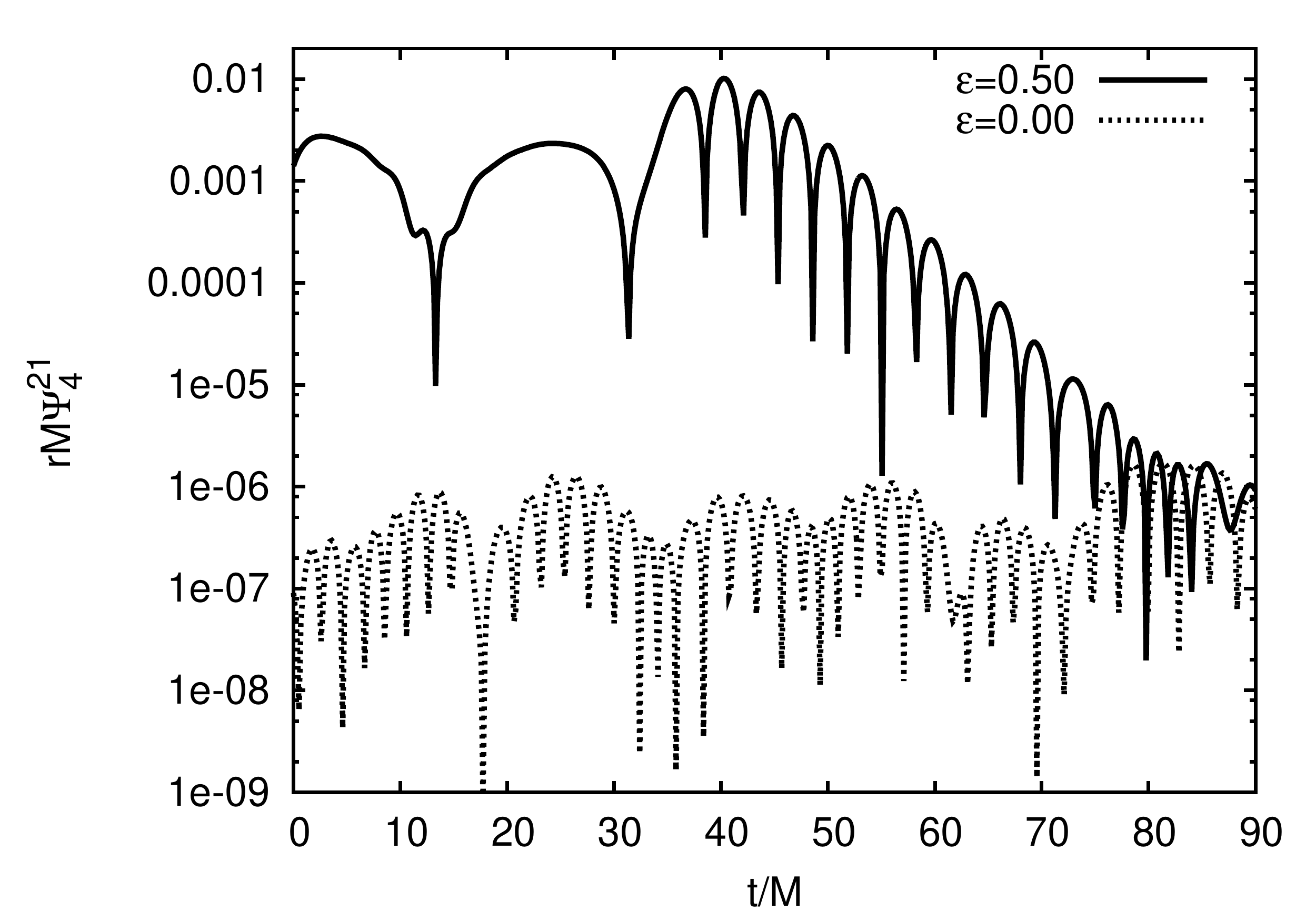}
    \includegraphics[width=.70\linewidth,angle=270]{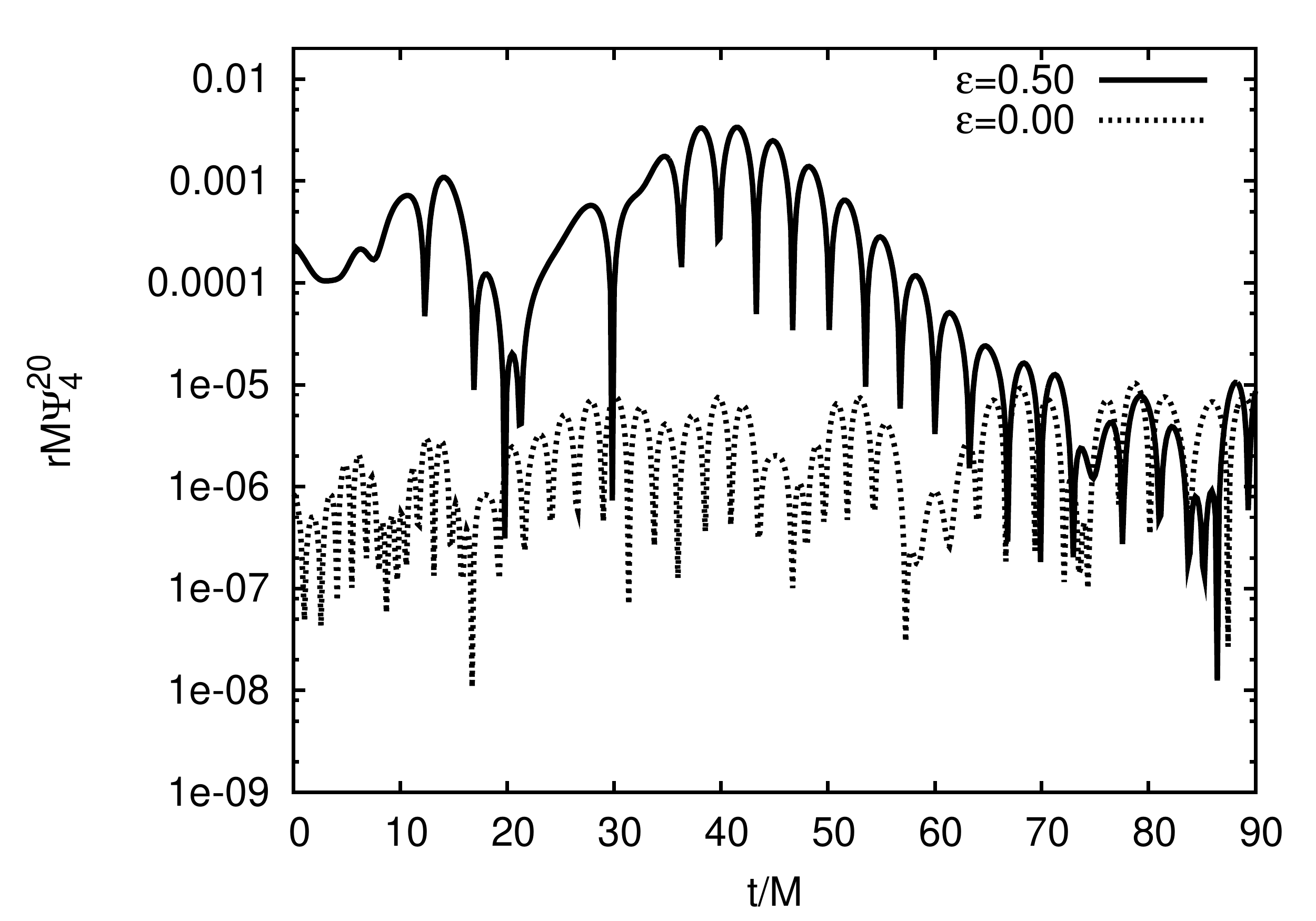}
}
\caption{From top to bottom, this figure shows the 22, 21, and 20 spin-weighted spherical harmonic modes of the Weyl scalar $\Psi_4$ for  $p = 0.0592$. The case $\epsilon = 0.5$ is depicted with a solid line and $\epsilon = 0$  with a dashed line.}
\label{fig:SCwfs}
\end{figure}

We have seen a strong indication of the predominance of an unstable spherical mode in the scalar field near black hole formation. It is then interesting to investigate the implications of this predominance on the emission of gravitational radiation. As noted, the initial scalar field configurations for $\epsilon \ne 0$  are $Y_{21}(\theta,\varphi)$ deformations of a spherical profile.  Thus, these deformation are in principle likely to trigger the emission of gravitational waves. 
Figure~\ref{fig:SCwfs} shows the 22, 21, and 20 spin-weighted spherical harmonic modes of the Weyl scalar $\Psi_4$ for the case $p = 0.0592$ with both, $\epsilon = 0.5$ (solid line) and $\epsilon = 0$ (dashed line). Not surprisingly, the 21 mode has the highest amplitude even though the scalar perturbation is spin weight 0 and the Weyl scalar has spin weight -2. The level of gravitational wave emission, $< 10^{-6}$, for $\epsilon = 0$ is below our numerical errors, thus consistent with the fact that gravitational waves cannot be emitted during spherically symmetric gravitational collapse. On the other hand, the waveforms from the $\epsilon = 0.5$ case show the characteristic shape of a burst followed by quasi-normal ringing found in a generic gravitational collapse. 
Interestingly, although the spherical mode in the scalar field dominates and yields a similar value of the critical exponent to that in spherical symmetry, the collapse retains a high enough level of non-spherical perturbations to trigger gravitational emission.  

As suggested by one of the referees, an interesting question to answer is the dependence of the gravitational radiation on $p-p^*$. Investigating this dependence during the quasi-normal ringing phase of the radiation is in particular illustrative.  As it is well known, the decay timescale and frequency of the quasi-normal ringing depend exclusively on the mass and spin of the black hole. We found that our collapse experiments yield effectively non-spinning black holes, with the largest dimensionless spin parameter of $O(10^{-4})$ for the $\epsilon = 1$ case.  Therefore, in the present study, the quasi-normal ringing is completely characterized by the mass of the black hole. As a consequence, given our result that $M_h/M \approx C(\epsilon) (p- p^*)^{0.37}$, the quasi-normal ringing observed in the $\Psi_4$ Weyl scalars depicted in Fig.~\ref{fig:SCwfs}  is fully determined by $p-p^*$ for a given $\epsilon$. 

%%%%%%%%%%%%%%%%%%%%%%%%%%%%%%%%%%%%%%%%%%%%%%%%%%%%%%%%%%%%%%%%%%%%%%%%%%%%
\section{Conclusions} 

Our study presents results from a 3D numerical relativity study of anisotropic scalar field collapse to investigate Choptuik's critical phenomena beyond axisymmetry. 
Specifically, the initial anisotropy in the scalar field we considered was proportional to the real part of the spherical harmonic $Y_{21}(\theta,\varphi)$. We found the typical black hole mass scaling of $M_h \propto (p-p^*)^\gamma$ expected in critical gravitational collapse. Independent of the strength of the anisotropy, our collapse simulations yielded scaling exponents   $\gamma \approx 0.37$, values very similar to those found in spherical symmetry. Our results support the dominance of an unstable spherical mode during the onset of black hole formation. We are currently investigating if these conclusions hold for other families of anisotropic initial data, in particular types of initial data that are not perturbations of spherically symmetric profiles.  Our results show also hints of  discrete self-similarity. However, we are not able to claim with  confidence that the observed periodicity is the result of critical collapse echoing because the solutions in our study are unfortunately far from the critical solution.

%%%%%%%%%%%%%%%%%%%%%%%%%%%%%%%%%%%%%%%%%%%%%%%%%%%%%
\begin{acknowledgements}
We thank Matt Choptuik for helpful discussions and comments.  Work supported by NSF grants 0653443, 0855892,
0914553, 0941417, 0903973, 0955825. Computations at
Teragrid  TG-PHY120016 and Georgia Tech FoRCE cluster.
JH gratefully acknowledges the NSF for financial support from Grants
PHY-1305730 and PHY-0969855.
\end{acknowledgements}


\begin{thebibliography}{10}
\providecommand{\url}[1]{{#1}}
\providecommand{\urlprefix}{URL }
\expandafter\ifx\csname urlstyle\endcsname\relax
  \providecommand{\doi}[1]{DOI~\discretionary{}{}{}#1}\else
  \providecommand{\doi}{DOI~\discretionary{}{}{}\begingroup
  \urlstyle{rm}\Url}\fi

\bibitem{1993PhRvL..70.2980A}
{Abrahams}, A.M., {Evans}, C.R.: {Critical behavior and scaling in vacuum
  axisymmetric gravitational collapse}.
\newblock Physical Review Letters \textbf{70}, 2980--2983 (1993).
\newblock \doi{10.1103/PhysRevLett.70.2980}

\bibitem{2006PhRvL..96k1102B}
{Baker}, J.G., {Centrella}, J., {Choi}, D.I., {Koppitz}, M., {van Meter}, J.:
  {Gravitational-Wave Extraction from an Inspiraling Configuration of Merging
  Black Holes}.
\newblock Physical Review Letters \textbf{96}(11), 111,102--+ (2006).
\newblock \doi{10.1103/PhysRevLett.96.111102}

\bibitem{Bode:2009mt}
Bode, T., Haas, R., Bogdanovic, T., Laguna, P., Shoemaker, D.: {Relativistic
  Mergers of Supermassive Black Holes and their Electromagnetic Signatures}.
\newblock Astrophys. J. \textbf{715}, 1117--1131 (2010).
\newblock \doi{10.1088/0004-637X/715/2/1117}

\bibitem{2006PhRvL..96k1101C}
{Campanelli}, M., {Lousto}, C.O., {Marronetti}, P., {Zlochower}, Y.: {Accurate
  Evolutions of Orbiting Black-Hole Binaries without Excision}.
\newblock Physical Review Letters \textbf{96}(11), 111,101--+ (2006).
\newblock \doi{10.1103/PhysRevLett.96.111101}

\bibitem{1993PhRvL..70....9C}
{Choptuik}, M.W.: {Universality and scaling in gravitational collapse of a
  massless scalar field}.
\newblock Physical Review Letters \textbf{70}, 9--12 (1993).
\newblock \doi{10.1103/PhysRevLett.70.9}

\bibitem{2003PhRvD..68d4007C}
{Choptuik}, M.W., {Hirschmann}, E.W., {Liebling}, S.L., {Pretorius}, F.:
  {Critical collapse of the massless scalar field in axisymmetry}.
\newblock \prd \textbf{68}(4), 044007 (2003).
\newblock \doi{10.1103/PhysRevD.68.044007}

\bibitem{2007LRR....10....5G}
{Gundlach}, C., {Mart{\'{\i}}n-Garc{\'{\i}}a}, J.M.: {Critical Phenomena in
  Gravitational Collapse}.
\newblock Living Reviews in Relativity \textbf{10}, 5 (2007)

\bibitem{2013PhRvD..88j3009H}
{Hilditch}, D., {Baumgarte}, T.W., {Weyhausen}, A., {Dietrich}, T.,
  {Br{\"u}gmann}, B., {Montero}, P.J., {M{\"u}ller}, E.: {Collapse of nonlinear
  gravitational waves in moving-puncture coordinates}.
\newblock \prd \textbf{88}(10), 103009 (2013).
\newblock \doi{10.1103/PhysRevD.88.103009}

\bibitem{Husa:2004ip}
Husa, S., Hinder, I., Lechner, C.: Kranc: a mathematica application to generate
  numerical codes for tensorial evolution equations.
\newblock Computer Physics Communications \textbf{174}, 983--1004 (2006)

\bibitem{2011arXiv1111.3344L}
{L{\"o}ffler}, F., {Faber}, J., {Bentivegna}, E., {Bode}, T., {Diener}, P.,
  {Haas}, R., {Hinder}, I., {Mundim}, B.C., {Ott}, C.D., {Schnetter}, E.,
  {Allen}, G., {Campanelli}, M., {Laguna}, P.: {The Einstein Toolkit: A
  Community Computational Infrastructure for Relativistic Astrophysics}.
\newblock ArXiv e-prints  (2011)

\bibitem{1999PhRvD..59f4031M}
{Mart{\'{\i}}n-Garc{\'{\i}}a}, J.M., {Gundlach}, C.: {All nonspherical
  perturbations of the Choptuik spacetime decay}.
\newblock \prd \textbf{59}(6), 064031 (1999).
\newblock \doi{10.1103/PhysRevD.59.064031}

\bibitem{2007PhRvD..76l4014O}
{Olabarrieta}, I.., {Ventrella}, J.F., {Choptuik}, M.W., {Unruh}, W.G.:
  {Critical behavior in the gravitational collapse of a scalar field with
  angular momentum in spherical symmetry}.
\newblock \prd \textbf{76}(12), 124014 (2007).
\newblock \doi{10.1103/PhysRevD.76.124014}

\bibitem{2011CQGra..28b5011S}
{Sorkin}, E.: {On critical collapse of gravitational waves}.
\newblock Classical and Quantum Gravity \textbf{28}(2), 025,011 (2011).
\newblock \doi{10.1088/0264-9381/28/2/025011}

\end{thebibliography}
\end{document}